\documentclass[prb,twocolumn,showpacs,a4paper]{revtex4}
\usepackage{amssymb}
\usepackage{amsmath}
\usepackage{epsfig}
\usepackage{epstopdf}
\usepackage{bbm}
\usepackage{graphicx}
\usepackage{enumerate}
\usepackage{hyperref}

\begin{document}

\title
{Antiferromagnetism and the gap of a Mott insulator: Results from analytic continuation of the self-energy}

\author{Xin Wang$^{(1)}$, Emanuel Gull$^{(1),(2)}$, Luca de' Medici$^{(3)}$, Massimo Capone$^{(4)}$, and Andrew J. Millis$^{(1)}$\\\vspace{.1in}\small{\sl $^{(1)}$Department of Physics, Columbia University, New York, New York 10027, USA\\
$^{(2)}$ETH Z\"urich, 8093 Z\"urich, Switzerland\\
$^{(3)}$Laboratoire de Physique des Solides, Universit\`e Paris-Sud, CNRS,
 UMR 8502,
F-91405 Orsay Cedex, France\\
$^{(4)}$SMC Center, CNR-INFM, and Dipartimento di Fisica, Universit\`a di
Roma ``La Sapienza'', Piazzale A. Moro 2, I-00185, Rome, Italy and
ISC-CNR, Via dei Taurini 18, I-00185, Rome, Italy}}

\date{\today}

\begin{abstract}
Direct analytic continuation of the self energy is used to determine
the effect of antiferromagnetic ordering on the spectral function
and optical conductivity of a Mott insulator. Comparison of several
methods shows that the most robust estimation of the gap value is
obtained by use of the real part of the continued self energy in the
quasiparticle equation within the single-site dynamical mean field
theory of the two dimensional square lattice Hubbard model, where for $U$ slightly
greater than the Mott critical value, antiferromagnetism increases
the gap by about 80\%.
\end{abstract}

\maketitle

\section{Introduction}

Quantum Monte-Carlo (QMC) evaluations of imaginary time path
integrals \cite{Blankenbecler1981, Hirsch1986, Werner2006a,
Werner2006b} have improved to the point where they constitute one of
the basic techniques of condensed matter physics. Even for fermionic
problem, where the sign problem precludes direct simulation, the
theoretical developments associated with single-site
\cite{Georges1996} and cluster \cite{Maier04, Kotliar06} Dynamical
Mean Field Theory (DMFT) have enabled a powerful approximate
solution in terms of a quantum impurity model which for not too
large clusters is sign-free or at least has a tractable sign
problem. However, while the QMC methods have proven to be very
powerful in the study of static expectation values, obtaining
dynamical information has remained challenging. The available
techniques are based either on an Exact Diagonalization (ED)
method,\cite{Caffarel94,Capone07} where the number of states which
contribute to a given response function is so small that level
spacings become an issue, or on the analytic continuation of
imaginary-time data,\cite{ Silver90,Gubernatis91,Jarrell1996} which
involves a host of other uncertainties.

\begin{figure}[t]
\begin{center}
\includegraphics[height=7.5cm, angle=-90]{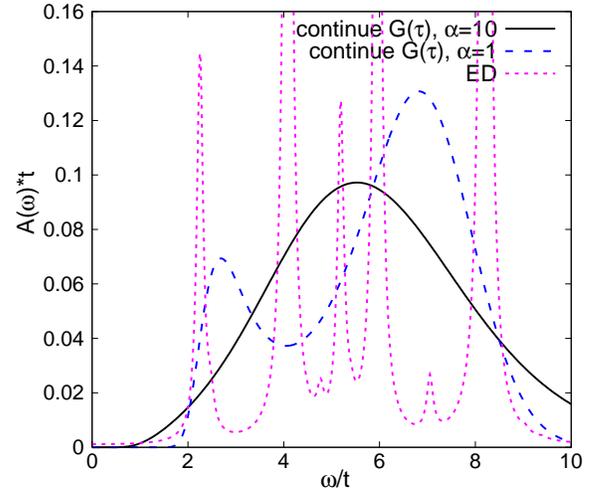}
\caption{Paramagnetic phase spectral function of the two dimensional
square lattice half-filled Hubbard model with  nearest neighbor
hopping and interaction parameter $U=12t$ computed using single site
DMFT with an ED impurity solver (dotted lines) or an imaginary time
hybridization expansion continuous time QMC impurity solver followed
by analytic continuation of the measured Green's function (solid and
dashed lines). Due to particle-hole symmetry $A(\omega)=A(-\omega)$
only positive frequency is shown. QMC is done at inverse temperature
$\beta t=10$, $\alpha$ is a parameter in analytic continuation
procedure which will be explained below.} \label{intro}
\end{center}
\end{figure}

The difficulties appear with particular force in the context of the
question of whether the high-$T_c$ cuprates are Mott insulating
materials.\cite{Imada98} Recent work has suggested that the value
of the gap and the form of the conductivity in the above-gap region
provide important insights into the physics of Mott and
charge-transfer insulators.\cite{Comanac08, deMedici2008, Weber08}
A question of particular interest is the change in gap value
associated with onset of antiferromagnetic order in a Mott
insulator. One recent paper argued in favor of negligible changes,
\cite{Weber08} while another argued for a large change.
\cite{Comanac08} However, determining with precision the gap value
in theoretical model of correlated material is not straightforward.
Fig.~\ref{intro} illustrates some of the uncertainties. It shows
three estimations of the local spectral function (many body density
of states) for a theoretical model (described more fully below) of a
Mott insulator: one obtained by an ED method and two obtained by
maximum entropy (MaxEnt)\cite{Jarrell1996} analytic continuation of
imaginary time QMC data. While the qualitative structure of the
three estimations appear consistent, there are significant
differences of detail, including a factor of two in the size of the
gap which makes it difficult to compare the theoretical results to
data.

In this paper we present a critical examination of  different
methods of determining the spectral function of a Mott insulator,
and apply the results to the question of the gap value,  spectral
function and optical conductivity in the paramagnetic and
antiferromagnetic phases of the two dimensional square lattice Hubbard model. We
study MaxEnt analytic continuation of the Green's function and of
the self-energy, and compare the results to ED calculations and to
direct thermodynamic evaluations of the gap. We argue that
continuation of the self energy provides the best method of
minimizing the broadening effect of MaxEnt procedure. The self
energy is also needed for computation of other response functions,
for example the optical conductivity. We establish that for $U$ near
the Mott critical value the onset of antiferromagnetism increases
the gap of the half filled square lattice Hubbard model by about
$3.4t$ relative to that of the paramagnetic case, thus increases the
gap by about 80\%.

The rest of this paper is organized as follows.  In section
\ref{Formalism} we define the model to be studied and the methods of
solution, discuss MaxEnt and in particular its application to the
self-energy. Section {\ref{Results}} presents a detailed analysis of
the results from analytic continuation and also summarizes the
interpretation of exact diagonalization data in light of these
results. Section \ref{Cond} discusses the optical conductivity and
section {\ref{Conclusion}} is a summary and conclusion.

\section{Formalism\label{Formalism}}
\subsection{Model}
We study the two dimensional Hubbard model defined by the Hamiltonian
\begin{equation}
H=\sum_{p,\sigma}\varepsilon_pc^\dagger_{p,\sigma}c_{p,\sigma}+U\sum_i\left(n_{i,\uparrow}-\frac{1}{2}\right)\left(n_{i,\downarrow}-\frac{1}{2}\right),
\label{H}
\end{equation}
with $\varepsilon_p=-2t\left(\cos p_x+\cos p_y\right)-\mu$. We
choose the chemical potential $\mu=0$ such that the electron density
$n=1$. We shall be interested in the imaginary part of the real-axis
electron Green's function.

To solve the model we employ the single-site dynamical mean field
approximation \cite{Georges1996} which makes the approximation that
the self energy is a function of frequency only
$\Sigma(p,\omega)\rightarrow\Sigma(\omega)$. On this assumption
Eq.~\eqref{H} may be mapped on to a quantum impurity model with
parameters determined by a self-consistency condition. The essential
computational task is to solve the quantum impurity model to obtain
the local Green's function $G_{\rm loc}$ and self energy
$\Sigma(\omega)$. We have used two methods: a  recently developed
hybridization expansion continuous-time QMC procedure
\cite{Werner2006a} and an ED method\cite{Caffarel94,Capone07}. The
ED method approximates physical response functions as a series of
poles. The QMC methods produces estimates of Green's functions and
self energies in imaginary time. One must then analytically continue
the QMC results to obtain physically relevant real frequency
quantities. To perform the analytic continuation we used the MaxEnt
methods pioneered in the condensed matter physics context by 
Gubernatis and co-workers {\sl et
al}.\cite{Silver90,Gubernatis91,Jarrell1996}

The qualitative behavior of the model is well understood.  For $n=1$
and any $U>0$, the ground state is antiferromagnetically ordered,
and the single particle spectrum is characterized by a gap,
$\Delta(U)$, which we would like to compute. For $U>U_{c2}\approx
12t$ there is a gap in the spectrum even if antiferromagnetism is
suppressed, and we are also interested in the value of this gap, and
in the behavior of the spectral functions for frequencies near the
gap edge.

To analyze the situation more precisely we note that in an insulator we expect $\mathrm{Im} G(\omega+i0^+)=0$ for $|\omega|<\Delta$. We may write the Green's function (in general a matrix) as
\begin{equation}
{\bf G}(p,\omega)=\left(\omega {\bf 1}-{\bf H}_0(p)-{\bf
\Sigma}(p,\omega)\right)^{-1}. \label{G}
\end{equation}
$\mathrm{Im} {\bf G}(p,\omega)\neq 0$ either when $\mathrm{Im}{\bf\Sigma}(p,\omega)\neq 0$ or, regardless of the value of $\mathrm{Im} \bf \Sigma$, if the ``quasiparticle equation''
\begin{equation}
\det \left[\omega  {\bf 1}-{\bf H}_0(p)-\mathrm{Re} {\bf \Sigma}(p,\omega)\right]=0
\label{qpeqn}
\end{equation}
is satisfied for some momentum $p$.

In the single-site DMFT the paramagnetic (PM) phase quasiparticle equation is:
\begin{equation}
\omega-\mathrm{Re}\Sigma(\omega)=-2t(\cos p_x+\cos p_y). \label{}
\end{equation}

In the antiferromagnetic (AFM) phase we have
\begin{equation}
{\bf G}^{-1}=\left(\begin{array}{cc}\omega-\Sigma_\uparrow(\omega) &
2t\left(\cos p_x+\cos p_y\right) \\2t\left(\cos p_x+\cos p_y\right)
& \omega-\Sigma_\downarrow(\omega)\end{array}\right), \label{g}
\end{equation}
and the quasiparticle equation is
\begin{equation}
\left(\omega-\mathrm{Re}\Sigma_\uparrow(\omega)\right)\left(\omega-\mathrm{Re}\Sigma_\downarrow(\omega)\right)=4t^2\left(\cos
p_x+\cos p_y\right)^2. \label{qpeqnafm}
\end{equation}

We may therefore define two gaps, $\Delta_{\mathrm{Im}\Sigma}$, the
lowest frequency at which $\mathrm{Im}\Sigma\neq 0$, and
$\Delta_{\rm qp}$, the lowest frequency at which the quasiparticle
equation is satisfied. If the interaction is non-vanishing, for
$\omega>\Delta_{\rm qp}$ phase space is available for a particle to
decay so that we expect $\Delta_{\rm
qp}\ge\Delta_{\mathrm{Im}\Sigma}$. Empirically we found $\Delta_{\rm
qp}\lesssim\Delta_{\mathrm{Im}\Sigma}$, suggesting that $\Delta_{\rm
qp}\simeq\Delta_{\mathrm{Im}\Sigma}$ for the Hubbard model.

We may also define a third gap $\Delta_\mu$ from the dependence of the particle
density $n$ on chemical potential $\mu$. This is given in terms of
the momentum-integrated spectral function $A_\sigma=-{\rm
Im}G_\sigma/\pi$ for spin $\sigma$ by
\begin{equation}
n(\mu)=\sum_\sigma \int d\omega
f\left(\omega-\mu\right)A_\sigma(\omega,\mu), \label{nofmu}
\end{equation}
where $f(\omega)=1/\left(\exp(\beta\omega)+1\right)$ is the Fermi
function. If the spectral function changes smoothly with $\mu$ then
for $\mu$ only slightly larger than $\Delta$ we would have, at $T=0$
\begin{equation}
n(\mu)=1+2\int_\Delta^\mu d\omega A(\omega,\mu=0)+{\cal
O}(\mu-\Delta)^2,
\end{equation}
so that $n(\mu)$ would change from 1 when $\mu=\Delta$. However, it
is known \cite{Fisher95,Kajueter95,Werner07} that in the PM Mott
insulating region of the single-site DMFT, the spectral function
changes nontrivially with chemical potential, introducing ``in-gap''
states so that $n$ begins to differ from unity at $\mu=\Delta_{\rm
PM}-A_0t$ with $A_0$ a number of order unity.  It is not known
whether this phenomenon occurs in the AFM phase.  The $\mu$ at which
$n$ begins to deviate from unity therefore provides a lower bound on
the gap in the insulating state.

\subsection{Analytic Continuation}
In practice QMC generates a numerical estimate $\bar{F}$ of a
function  $F(\tau)$ ($F(i\omega_n)$) defined on imaginary time (or Matsubara frequency). For fermionic correlators $F$ is related
to a spectral function $A(\omega)$ by:
\begin{equation}
F(\tau)=\int_{-\infty}^{\infty}d\omega\frac{e^{-\tau\omega}}{1+e^{-\beta\omega}}A(\omega)
\label{GA}
\end{equation}
with $\beta=1/T$ the inverse temperature.

Analytical continuation is the inversion of Eq.~\eqref{GA} to
determine $A$ given $F$. Unfortunately, the matrix defined by
$\exp(-\tau\omega)/\left(\exp(-\beta\omega)+1\right)$ is extremely
poorly conditioned, with many relatively small eigenvalues which, on
inversion, greatly amplify any errors in $\bar{F}$ (i.e. differences between QMC estimate $\bar{F}$ and true value $F$)
leading to highly unreliable estimates of $A$. While various
attempts have been made to avoid this problem, the most widely used one
is the MaxEnt method.\cite{
Silver90,Gubernatis91,Jarrell1996} MaxEnt is based on defining $A$ as the
function which extremizes a cost functional $Q[\{A\}]$ which
is the sum of entropy-like ($S$) and energy-like ($L$) terms:
\begin{equation}
Q[\{A\}]=\alpha S[\{A\}]-L[\{A\}]. \label{QSL}
\end{equation}
In Eq.~\eqref{QSL}, $\alpha$ is a temperature-like quantity that
controls the competition between $S$ and $L$.  The energy-like term
is defined in terms of the mean square misfit between the proposed
spectrum computed from Eq.~\eqref{GA} and the QMC data $\bar{
F}$, which in the matrix form is:
\begin{equation}
L=\frac{1}{2}(\bar{\bf F}-{\bf KA})^T{\bf C}^{-1}(\bar{\bf F}-{\bf
KA}).\label{}
\end{equation}
Here ${\bf C}$ is a correlation matrix which represents the uncertainties (statistical and systematic) in the computation:
\begin{equation}
C_{ij}=\langle\delta F(\tau_i)\delta F(\tau_j)\rangle.\label{}
\end{equation}

The crucial part of the method is the entropy-like term, which is defined in terms of a model function $m(\omega)$ as
\begin{equation}
S=\int d\omega
\left[A(\omega)-m(\omega)-A(\omega)\ln\frac{A(\omega)}{m(\omega)}\right].
\label{}
\end{equation}
The model function is chosen to encapsulate prior information about
the function $A$: in the problems of physical relevance this
typically includes positivity and a known normalization. We use a
Gaussian model function
$1/(\sqrt{2\pi}\sigma)\exp{(-x^2/(2\sigma^2))}$ with $\sigma=5$ and
we have checked that the results do not depend on the width of this
Gaussian. To perform the minimization we use the algorithm of
Ref.~\onlinecite{Jarrell1996}. We generate spectra using a broad
range of $\alpha$ and select the spectrum corresponding to the most
probable $\alpha$, according to Ref.~\onlinecite{Jarrell1996}, by
calculating the posterior probability of $\alpha$ at a given ${\bar
G}$ $P(\alpha|{\bar G})$ which up to a normalization factor is:
\begin{equation}
P(\alpha|{\bar
G})=\prod_i\left(\frac{\alpha}{\alpha+\lambda_i}\right)^\frac{1}{2}\frac{e^{Q(\hat{A})}}{\alpha},
\end{equation}
where $\hat{A}$ is the resulting spectrum and $\lambda_i$ are the
eigenvalues of $\{A^{1/2}\}\nabla\nabla L|_{\hat{A}}\{A^{1/2}\}$.
Here $\{A^{1/2}\}$ means a matrix with elements
$\sqrt{A_i}\delta_{ij}$.

Note that uncertainties of error bar estimation from binning of 
Monte Carlo data and neglect of off diagonal correlation matrix 
elements (explained below) may introduce errors in selected $\alpha$ from $P(\alpha|G)$.
 In the following context we shall show that a reliable determination 
 of gap size does not change with reasonable variance of $\alpha$.

\subsection{Self energy}

The feature of our work is the direct continuation of the self
energy which is related to the full Green's function $G$ and the
noninteracting Green's function $G_0$ by
\begin{equation}
\Sigma=G_0^{-1}-G^{-1}. \label{sigdef}
\end{equation}

In the Hubbard model,  the self energy has the following asymptotic
behavior:\cite{Potthoff97}
\begin{equation}
\Sigma_\sigma(i\omega_n)=U\langle n_{-\sigma}\rangle+U^2\langle
n_{-\sigma}\rangle(1-\langle
n_{-\sigma}\rangle)\frac{1}{i\omega_n}+O(\frac{1}{(i\omega_n)^2}).
\label{asymbeh}
\end{equation}

If the Hartree term $U\langle n_{-\sigma}\rangle$ is subtracted the
remaining is just like the Green's function with a different
normalization $U^2\langle n_{-\sigma}\rangle(1-\langle
n_{-\sigma}\rangle)$. We also introduce the cutoff frequency
$\omega^*$ to avoid the error in high frequencies and replace the
self energy with frequencies above $\omega^*$ by its known
asymptotic behavior \eqref{asymbeh}. Thus frequencies above
$\omega^*$ will not be included in the MaxEnt procedure and their
value should be correctly reproduced if we satisfy the normalization
condition. We have also verified that including points at $\omega>\omega^*$ does not change our result since (as discussed below) the
${\rm Im}\Sigma$ at higher Matsubara frequencies have a much larger
error estimation thus contribute much smaller weight in calculating
$L$.



It is important to properly treat the noise correlations and
encode them in the correlation matrix. In our work we
do not consider errors in $G_0$. Thus if $G$ has an average value ${\bar
G}$ and a measurement error $\delta G$ then by expanding
\begin{equation}
\delta \Sigma(i\omega_n)={\bar G}^{-2}(i\omega_n)\delta
G(i\omega_n)+{\bar G}^{-3}(i\omega_n)\delta G(i\omega_n)^2 +\cdots
\label{deltaGdef}
\end{equation}
Therefore $\Sigma$ has a possibly non-zero shift:
\begin{equation}
\langle\delta\Sigma(i\omega_n)\rangle=\bar
G^{-3}(i\omega_n)\langle\delta G(i\omega_n)^2\rangle+\cdots
\label{sigshift}
\end{equation}
and a correlator
\begin{equation}
\begin{split}
&\langle\delta
\Sigma(i\omega_n)\delta\Sigma^*(i\omega_m)\rangle\\&={\bar
G^{-2}}(i\omega_n){\bar G^{*-2}}(i\omega_m)\langle\delta
G(i\omega_n)\delta G^*(i\omega_m)\rangle+\cdots
\end{split}
\label{sigcorr}
\end{equation}
Structure in $G$ means that there are important correlations in
$\Sigma$; in particular, because at large $\omega$, $G\sim
1/\omega$, the high frequency fluctuations in $\Sigma$ are large.

We estimate $G(\tau)$ from continuous time measurements binned into
uniformly discretized bins of width $\Delta\tau$. Suppose that in
imaginary time the fluctuations in $G$ are $\delta$-correlated and
independent of time (this may always be ensured by an appropriate
measurement process)
\begin{equation}
\langle\delta G(\tau_i) \delta G(\tau_j)\rangle=g^2\delta_{ij}.
\label{gcorrtau}
\end{equation}
Then (assuming $\tau$ are evenly discretized on $[0,\beta)$ with size $\Delta\tau$)
\begin{equation}
\langle\delta G(i\omega_n) \delta G^*(i\omega_m)\rangle=\beta
g^2(\Delta\tau)\delta_{\omega_n,\omega_m}. \label{gcorrw}
\end{equation}
Note $g^2\Delta\tau$ is expected to be independent of bin size
$\Delta\tau$. Then Eq.~\eqref{sigshift} gives zero and
Eq.~\eqref{sigcorr} gives
\begin{equation}
\begin{split}
\langle&\delta \Sigma(i\omega_n) \delta
\Sigma^*(i\omega_m)\rangle\\&={\bar G^{-2}}(i\omega_n){\bar
G^{*-2}}(i\omega_m)\beta g^2(\Delta\tau)\delta_{\omega_n,\omega_m},
\end{split}
\label{sigcorr1}
\end{equation}
which is the correlation matrix we need.
Observe that this means that in this case the orthogonal transformation which diagonalizes the covariance matrix is the transformation to Matsubara frequencies.

\subsection{Continuing $G(\tau)$}

\begin{figure}
    \centering
    \includegraphics[height=7.5cm, angle=-90]{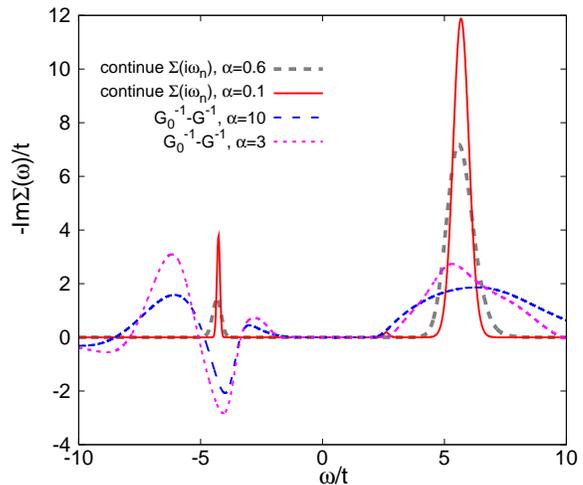}
    \caption{Comparison of minority spin $\mathrm{Im}\Sigma(\omega)$ computed by subtracting continued $G_0^{-1}$ and $G^{-1}$(long and short dashed lines), and continuing $\Sigma$ (thick dashed and the solid lines) for $\beta t=10$, $U=12t$ half-filled square lattice in the antiferromagnetic phase. We see that the $G_0^{-1}-G^{-1}$ one does not preserve positive definiteness.}
    \label{SEAFMGtau}
\end{figure}

The self energy may alternatively be computed by first continuing $G_0$ and $G$, inverting the continued functions, and subtracting.  Fig.~\ref{SEAFMGtau} compares the imaginary part of self-energy computed in this way to the result obtained by continuing $\Sigma(i\omega_n)$. We present two $\alpha$ values for each computation, one chosen to be near the peak of $P(\alpha|{\bar G})$ ($\alpha=3$) and one at a somewhat larger $\alpha$ ($\alpha=10$).

The quantity $G_0^{-1}-G^{-1}$ is not  guaranteed to be positive definite and we see that an unphysical sign indeed occurs. The difficulties are that $G$ and $G_0$ are small at high frequency so that errors in the MaxEnt procedure are amplified on inversion.  Errors in the position of the gap edge are also amplified in $\Sigma$. Finally, the calculated structures in $\Sigma$ are too broad. Another deficiency is seen in the PM insulating phase (not shown here) where $G_0^{-1}-G^{-1}$ fails to
correctly represent the pole at the chemical potential which is
known to exist in $\Sigma$.

\section{Results\label{Results}}

\subsection{Method}

The DMFT calculation was performed with the hybridization expansion
continuous-time QMC solver \cite{Werner2006a}. Typically more than
$10^9$ Monte Carlo steps are made in each DMFT iteration, which
usually takes around one hour CPU time on a cluster with 40 2GHz
processors. Special attention must be paid in the paramagnetic
insulating phase: in order to resolve the pole in $\Sigma(\omega)$
one need a real frequency grid which has very high resolution in the
vicinity of the chemical potential.

We use $\alpha$ values that range several orders of magnitude to do analytic continuation. For each given $\alpha$ we calculate $P(\alpha|{\bar G})$ at the convergence, use Kramers-Kronig relation to get ${\rm Re}\Sigma(\omega)$ from computed ${\rm Im}\Sigma(\omega)$, and do a momentum integral of Eq.\eqref{G} to get the Green's function.

\subsection{PM phase}

\begin{figure}
    \centering
    \includegraphics[height=7.5cm, angle=-90]{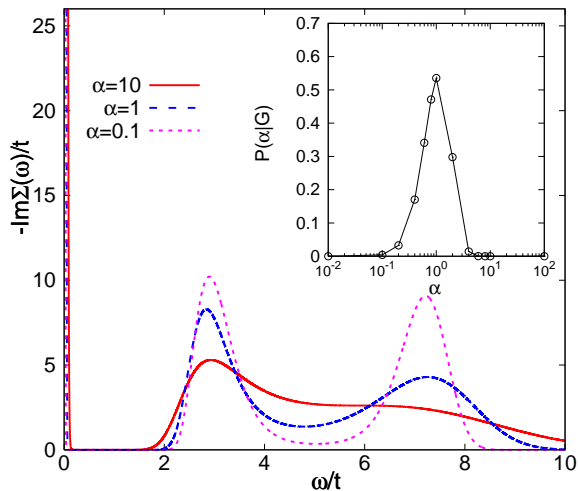}
    \caption{Main panel: Imaginary part of PM self-energy for $\beta t=10$, $U=12t$, $\mu=0$ with several $\alpha$'s. Inset: the posterior probability $P(\alpha|{\bar G})$ as a function of $\alpha$. Note the log scale of $\alpha$. At $\alpha=10$ (above the maximum of $P(\alpha|{\bar G})$) the result is very smooth. At $\alpha=1$ (at the maximum of $P(\alpha|{\bar G})$) more detailed features appear, which become more pronounced for $\alpha=0.1$ (below the maximum of $P(\alpha|{\bar G})$).}
    \label{ImSEPM}
\end{figure}

The main panel of Fig.~\ref{ImSEPM} shows the imaginary part of
continued self-energies calculated with three $\alpha$ value for the
paramagnetic insulating phase of the two dimensional square lattice
Hubbard model. We see a clear pole near the chemical potential but
the detailed structures in side bands $2t<\omega<8t$ vary. The inset
shows $P(\alpha|{\bar G})$ with a maximum at $\alpha=1$. For
$\alpha=10$ (above the maximum of $P(\alpha|{\bar G})$), ${\rm
Im}\Sigma$ is smooth because of the regularization from model
function. For $\alpha=1$ (at the maximum of $P(\alpha|{\bar G})$),
more structure is observed. For $\alpha=0.1$ (below the maximum of
$P(\alpha|{\bar G})$) these detailed features are more pronounced.
The differences in the curves give some idea of the uncertainties in
the process. The maximum in posterior probability identifies the
$\alpha=1$ curve as the preferred continuation.

\begin{figure}
    \centering
    \includegraphics[height=7.5cm, angle=-90]{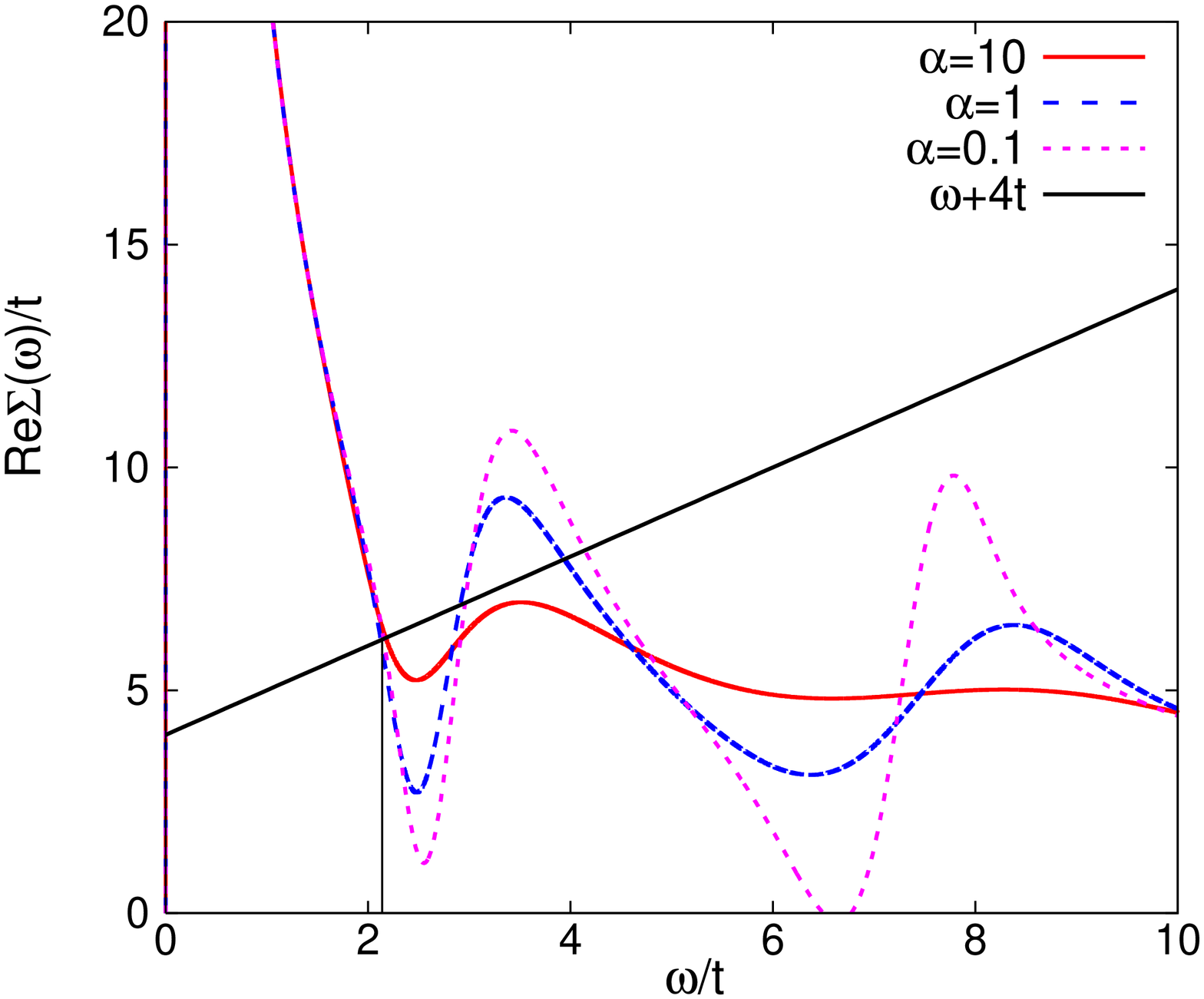}
    \includegraphics[height=7.5cm, angle=-90]{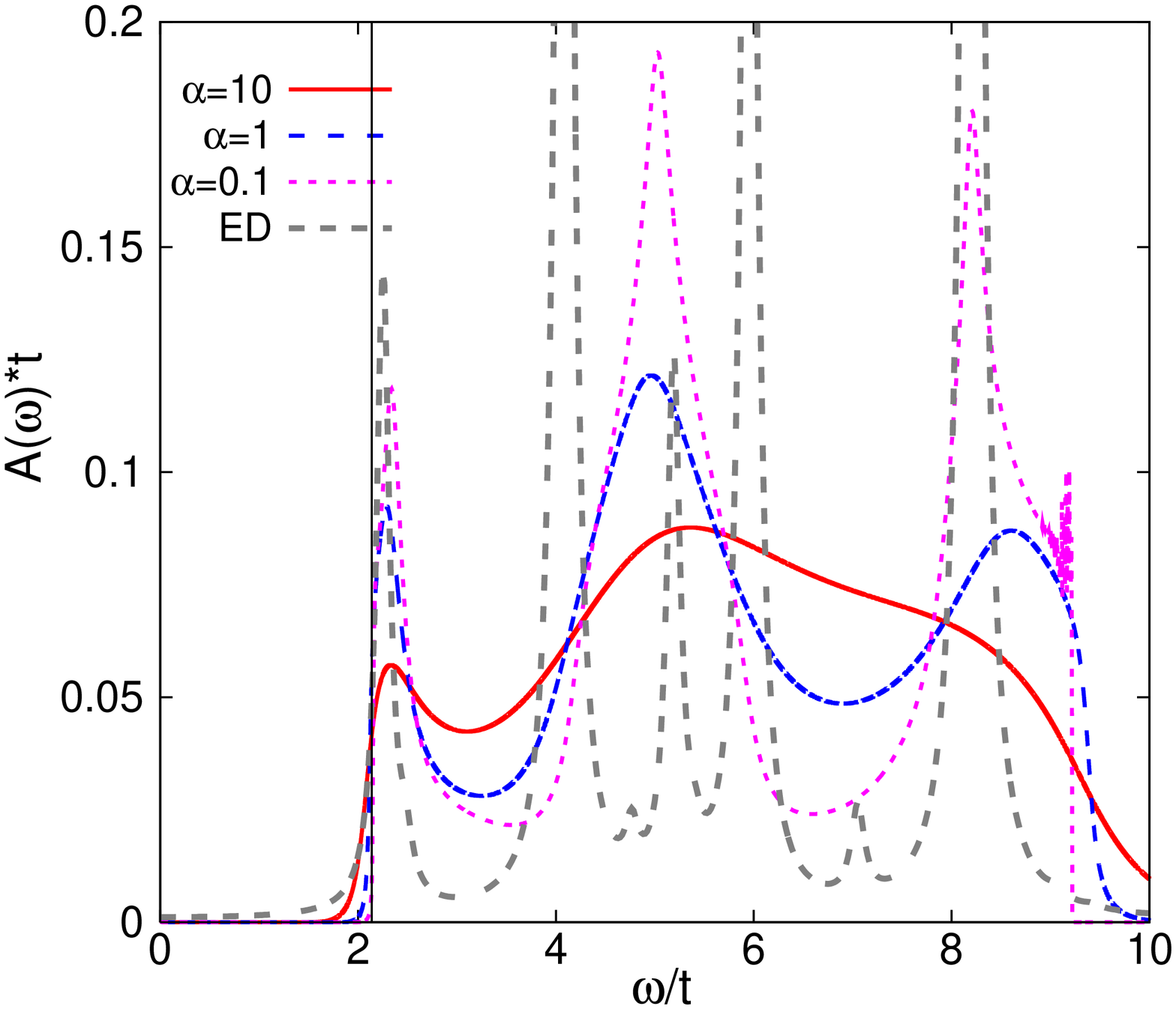}
    \caption{Upper panel: Real part of PM self-energies for $\beta t=10$, $U=12t$, $\mu=0$  for  several values of $\alpha$, along with quasiparticle equation $\mathrm{Re}\Sigma(\omega)=\omega+4t$. The quasiparticle equation has a minimum positive solution at around $\omega=2.1t$ Lower panel: Spectral function constructed from the continued self energy along with the result from ED. Vertical lines at $\omega=2.1t$ are drawn as eye guide. We see that all curves are consistent with the estimated half gap size $\omega=2.1t$.}
    \label{SEAomegaEDPM}
\end{figure}

The upper panel of Fig.~\ref{SEAomegaEDPM} shows the real part of
continued self energies. The $\alpha$ values are the same as those
in Fig.~\ref{ImSEPM}. The pole at $\omega=0$ in $\rm{Im}\Sigma$
implies $\rm{Re}\Sigma\sim1/\omega$ at low frequencies. The crossing
point between $\mathrm{Re}\Sigma$ and $\omega+4t$ curves gives the
minimal positive solution to the quasiparticle equation, which gives
an estimated gap size around $\omega=2.1t$. Turning to the lower
panel which is the spectral function plotted at the same
$\omega$-scale, we see that the estimate size of  $2.1t$ is
consistent with all four curves: ED puts its first peak slightly
above $\omega=2.1t$, and the non-zero structure  of QMC curves below
$\omega=2.1t$ could be safely considered as a result of broadening
in MaxEnt procedure. The interesting fact is that the estimate of
gap size is robust against a reasonable variation of $\alpha$ which
provides an indication that the gap estimate is reliable.

\begin{figure}
    \centering
    \includegraphics[height=7.5cm, angle=-90]{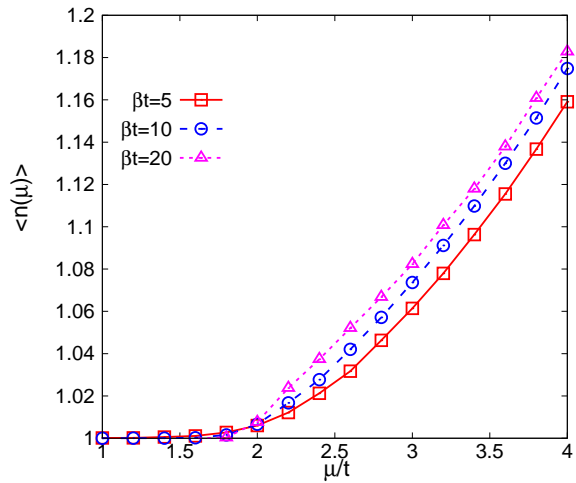}
    \caption{Averaged total electron density as a function of chemical potential $\langle n(\mu)\rangle$ for paramagnetic $U=12t$ square lattice at inverse temperatures $\beta t=5,10,20$. We see that $\Delta_\mu=1.8t$.}
    \label{nvsmuPM}
\end{figure}

Fig.~\ref{nvsmuPM} shows the averaged total electron density as a function of
chemical potential at $U=12t$ and various different temperatures.
This does not need analytic continuation thus provides an
independent check of the MaxEnt results. If we assume $A(\omega)$
changes slowly with temperature, then thermal fluctuation gives
$\langle n(T)\rangle=\langle n(T=0)\rangle+AT^2$ with $A$ a positive
number if $\mu$ is close to the lower edge of the upper Hubbard band.
However, on the contrary, $\langle n\rangle$ increases as
temperature is decreased in Fig.~\ref{nvsmuPM}. This suggests
existence of ``in-gap'' states which increase rapidly as temperature
is reduced. Moreover, for the lowest available temperature $\beta
t=20$ curve we see an almost linear dependence close to the band
edge ($2t<\omega<2.5t$) which is also a result of ``in-gap'' states.
Extrapolation of the $\langle n(\mu)\rangle$ curve gives
$\Delta_\mu=1.8t$ which, considering the presence of ``in-gap''
states, is consistent with the discussion in section~\ref{Formalism}
that $\Delta_{\rm qp}-\Delta_\mu=0.3t$.

\subsection{AFM phase}

\begin{figure}
    \centering
    \includegraphics[height=7.5cm, angle=-90]{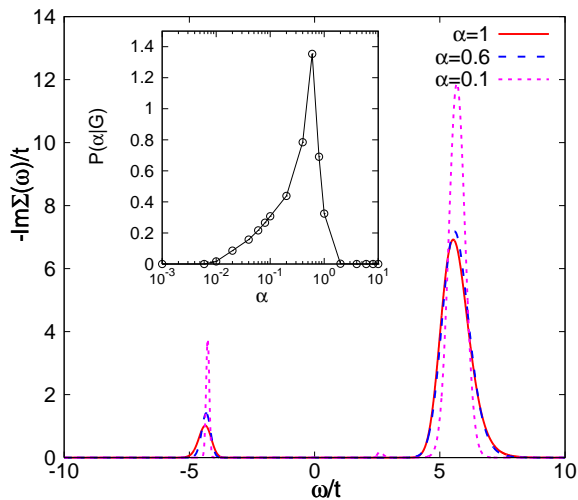}
    \caption{Main panel: Imaginary part of AFM minority spin self-energy for $\beta t=10$, $U=12t$,
    $\mu=0$ with several values of $\alpha$. Inset: $P(\alpha|{\bar G})$ versus $\alpha$ curve. We again see that the continued self-energy has sharper features at $\alpha=0.1$ (below the maximum of $P(\alpha|{\bar G})$) than at $\alpha=0.6,1$ (at or above the maximum of $P(\alpha|{\bar G})$).}
    \label{ImSEAFM}
\end{figure}

The main panel of Fig.~\ref{ImSEAFM} is the continued imaginary part
of $\Sigma(\omega)$. The inset shows the corresponding
$P(\alpha|{\bar G})$ curve. We see that the $\alpha=0.1$ curve is
sharper than the curves for $\alpha=0.6,1$. This is very similar with what
we found in PM case.

\begin{figure}
    \centering
{\includegraphics[height=7.5cm, angle=-90]{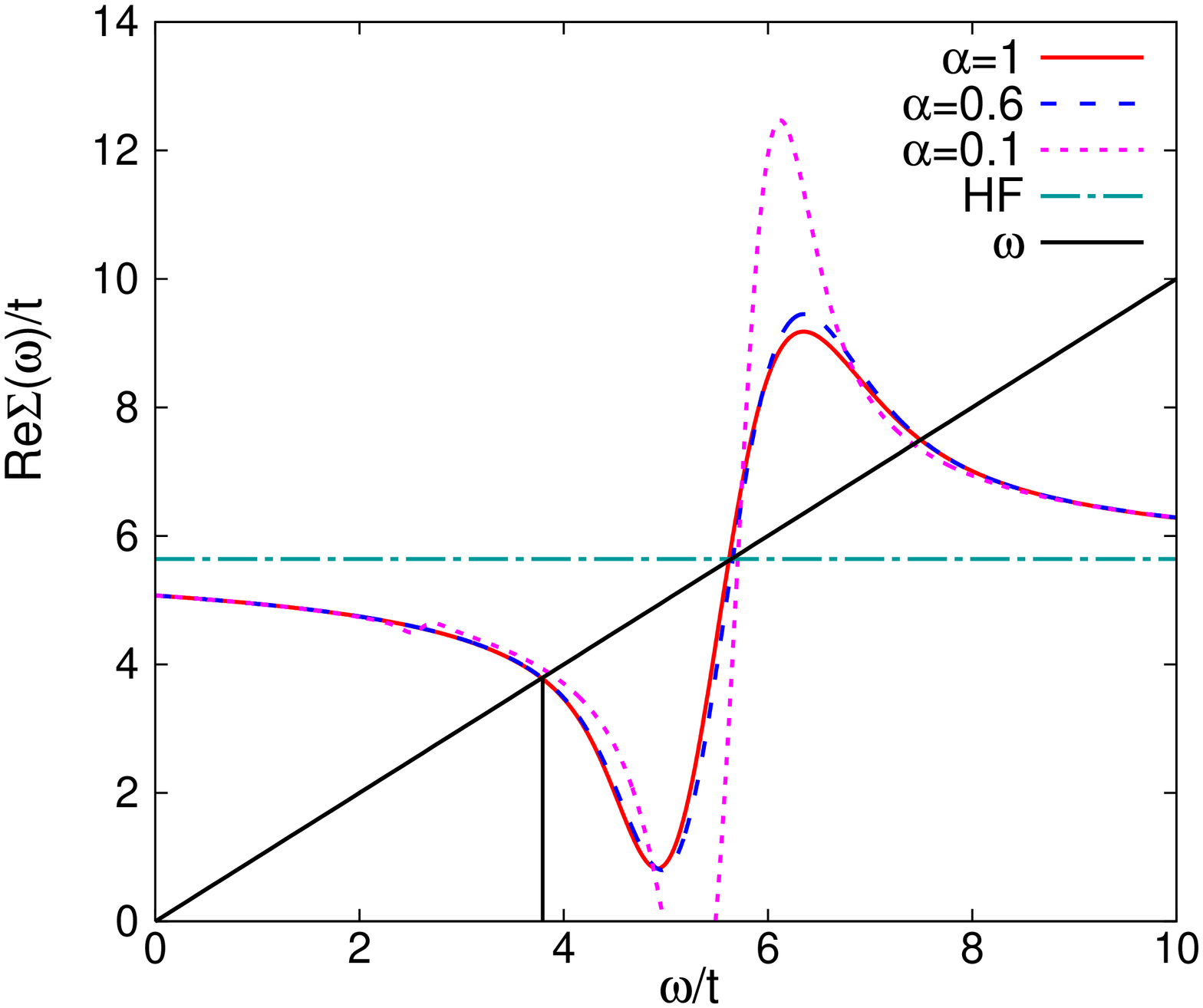}}
{\includegraphics[height=7.5cm, angle=-90]{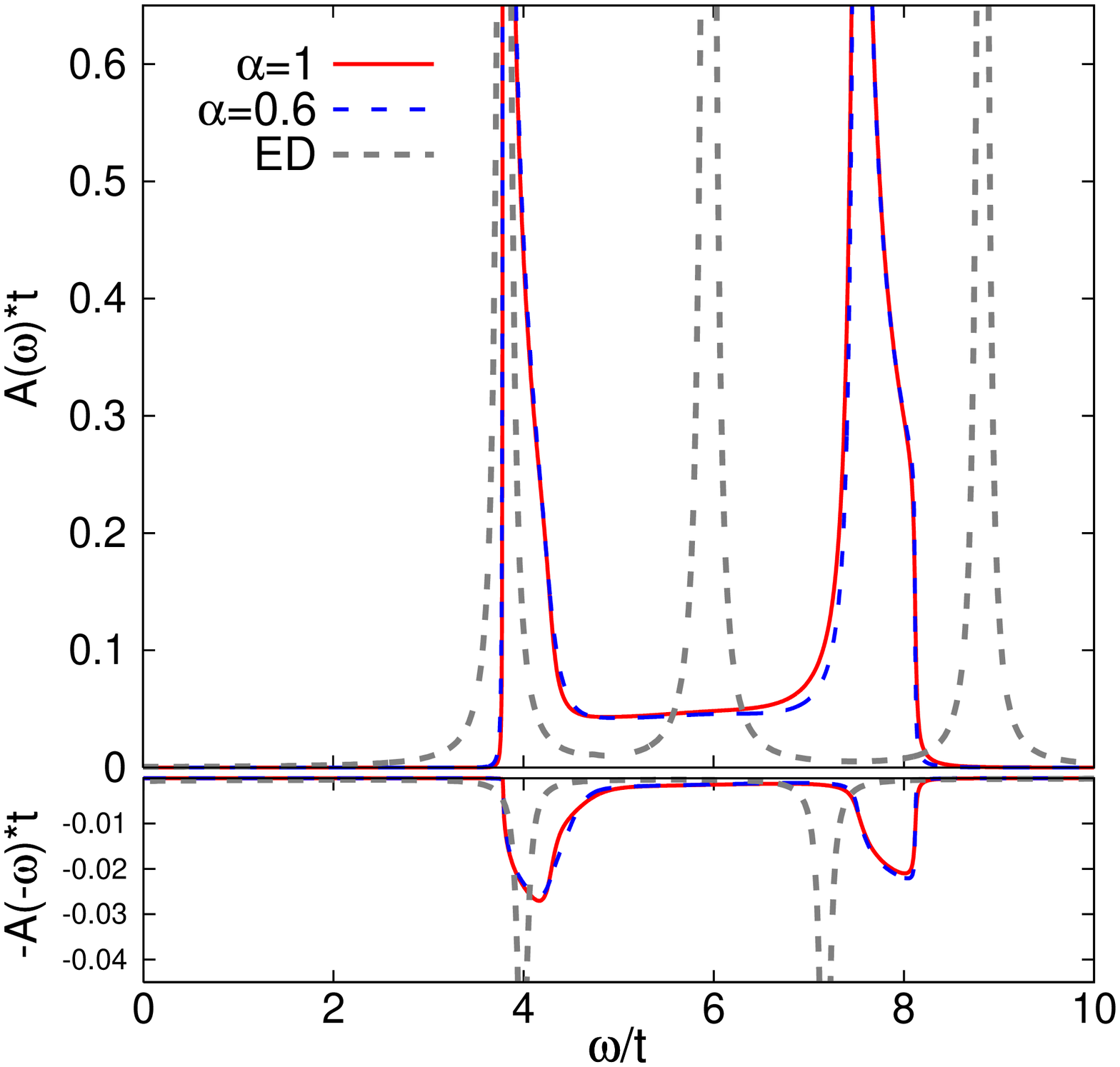}}
    \caption{Upper panel: Real part of AFM self-energy for $\beta t=10$, $U=12t$, $\mu=0$ with several values of $\alpha$, and the quasiparticle equation $\mathrm{Re}\Sigma(\omega)=\omega$. The Hartree-Fock mean field prediction (HF) is also shown. The quasiparticle equation has a solution at $\omega=3.8t$. A vertical line is drawn at $\omega=3.8t$ as an eye guide. Lower panel: Spectral function constructed from the continued self energy along with ED result. Due to the particle-hole symmetry $A_{\rm majority}(\omega)=A_{\rm minority}(-\omega)$ only minority spin is shown. The removal peak of spectral function has been reflected to positive frequency and shown in the bottom of the lower panel. The estimate of gap edge $\omega=3.8t$ is consistent with the reconstructed spectral function.}
    \label{SEAomegaEDAFM}
\end{figure}

The upper panel of Fig.~\ref{SEAomegaEDAFM} shows the minimum
positive solution to the quasiparticle equation. We get a half gap
size around $3.8t$. This is also robust against changing $\alpha$
with a variance as small as $0.1t$. 
The dash-dotted line shows the Hartree-Fock mean field solution. 
The averaged magnetization produced by Hartree-Fock $\langle m\rangle=0.94$ 
agrees with the QMC value, 
but it predicts a gap size $5.6t$ much larger than the QMC value. 

The lower panel of
Fig.~\ref{SEAomegaEDAFM} shows the minority spin AFM spectral
function constructed from the continued self energy. ($A_{\rm
majority}(\omega)=A_{\rm minority}(-\omega)$) The behavior at the
gap edge is very sharp. The sharpness comes from the combined
effect of Fermi surface nesting and mass renormalization. The half
gap size is consistent with $\Delta_{\rm qp}=3.8t$. As in PM
phase,  ED also put its first peak at the gap edge.
At higher frequency there is a clear difference between ED and QMC 
calculations: QMC has one additional peak at around $7.5t$ while 
ED has two peaks at $6t$ and $9t$ and in particular, the estimate 
of the upper band edge is different. As noted in section I, ED 
and QMC+MaxEnt make different approximations and thus produces 
different self energies (especially at high frequencies), 
which translates to a difference in estimate of upper band edge mathematically. However, at current stage we do not have definite statements about its physical origin.

\begin{figure}
    \centering
    \includegraphics[height=7.5cm, angle=-90]{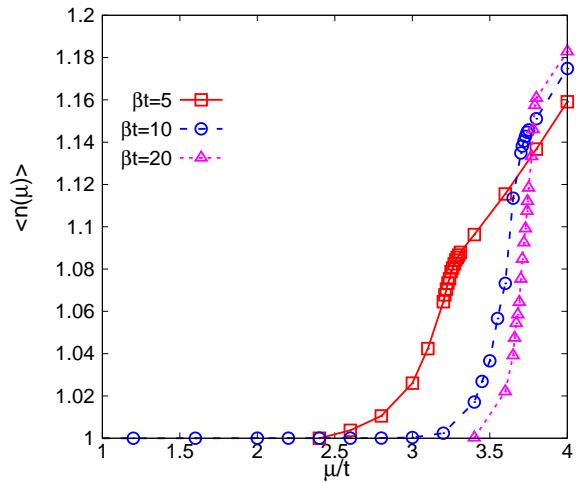}
    \caption{Averaged total electron density as a function of chemical potential $\langle n(\mu)\rangle$
for the Hubbard model on a square lattice using a self consistency
condition that allows for AFM order, at $U/t=12$ and inverse
temperature $\beta t=5$, 10, and 20.}
    \label{nvsmuAFM}
\end{figure}

Fig.~\ref{nvsmuAFM} shows the averaged total electron density as function of
$\mu$ when AFM order is allowed. As in PM phase, this provides an
independent check of MaxEnt results. For the temperature studied
($\beta t=10$), $\Delta_\mu\simeq3.6t\lesssim\Delta_{\rm qp}$. This
is again consistent with the discussion in Section~\ref{Formalism}.
We also see that the gap will get larger when temperature is
reduced. Within our precision we cannot distinguish whether
``in-gap'' states exist in this case. We see that varying the
chemical potential in the insulating phase leads at low $T$ to a
very sharp transition (visible as a slope discontinuity in the
$n(\mu)$ curves) between a paramagnetic metal phase and an
antiferromagnetic insulating phase.  Even at our lowest temperature
$\beta t=20$ the $\langle n(\mu)\rangle$ curve is apparently
continuous and is associated with a rapid (but also apparently
continuous) change of staggered magnetization (not shown here).
Whether the transition becomes first order as $T\rightarrow0$
remains to be studied. On increasing $\mu$ from 0 the transition
occurs at $\Delta_\mu\lesssim\Delta_{\rm qp}$. (At $\beta t=10$,
$\Delta_\mu\simeq3.6t$ and $\Delta_{\rm qp}=3.8t$.)

\section{Optical conductivity\label{Cond}}

\begin{figure}
    \centering
    \includegraphics[height=7.5cm, angle=-90]{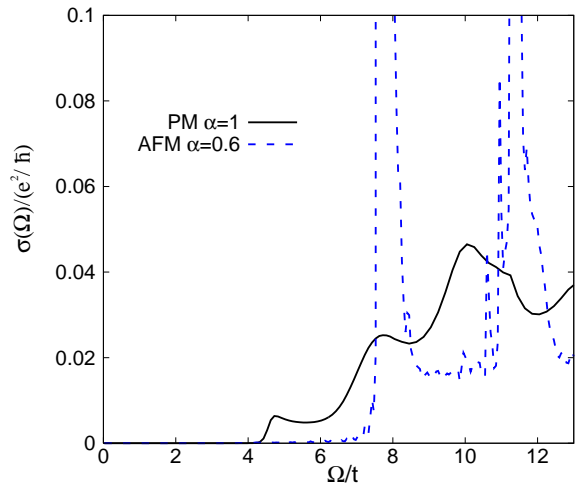}
    \caption{Optical conductivities constructed from analytically continued QMC data for
    half-filled Hubbard model on a square lattice at $U=12t$ and inverse temperature $\beta t=10$. The PM curve has a gap at $\Omega\simeq4.2t$ and the AFM curve has a sharp gap edge at $\Omega\simeq7.6t$.}
    \label{cond}
\end{figure}

The  optical conductivity can be computed using  the Kubo formula
and the minimal coupling ansatz  ${\bf p} \rightarrow {\bf p}-{\bf
A}$. The dissipative part of the conductivity is then\cite{Millis05}
\begin{equation}
\begin{split}
\sigma(\Omega)=\frac{2e^2}{\hbar}\int_{-\infty}^\infty \frac{d\omega}{\pi}\int \frac{d^2p}{(2\pi)^2}\frac{f(\omega)-f(\omega+\Omega)}{\Omega}\\
\times{\rm Tr}\left[{\bf j}(p){\rm Im}{\bf G}(\omega+\Omega,p){\bf
j}(p){\rm Im}{\bf G}(\omega,p)\right],
\end{split}
\label{sigmamatrix}
\end{equation}
where the current operator is ${\bf j}=\delta{\bf H}/\delta p_x$.

Fig.~\ref{cond} compares the paramagnetic and antiferromagnetic
phase optical conductivities. We see that PM optical conductivity
has a gap $\simeq4.2t$ and a relatively soft edge, while in AFM
phase it has a gap of $\simeq7.6t$ and a sharp edge. This is a consequence of the large change in gap due to antiferromagnetism. The
high-$T_c$ cuprates are believed to be described by $t\approx0.38$eV
(this value is for example the average of the even and odd parity
values quoted in the table in section 7 of
Ref.~\onlinecite{Andersen95}). Our result would imply that if $U$ in
the cuprates were of the order of $U_{c2}$ the optical gap would be
about  2.9eV, rather larger than the observed
$2\Delta\approx1.8$eV.

\section{Conclusion\label{Conclusion}}

To conclude, we have presented a method to find the gap size from
QMC DMFT calculations. We first continue the measured
$\Sigma(i\omega_n)$ to $\Sigma(\omega)$, using MaxEnt with the
correctly estimated correlation matrix, and select $\alpha$
from the peak in the posterior probability $P(\alpha|{\bar G})$. We
then plot $\mathrm{Re}\Sigma(\omega)$ and find the lowest positive
solutions to the quasiparticle equation. Curves corresponding to
different values of $\alpha$ may give slightly different estimates,
but this variation has been found to be small. We find that within
our numerical accuracy the gap edge is defined by the quasiparticle
equation Eq.~\eqref{qpeqn} so (at least within the single-site DMFT)
there are no ``in-gap'' excited states arising from e.g. an
excitonic binding between a particle and a spin wave. In
the paramagnetic phase doping produces ``in-gap'' states; we have
established that the shift is about $0.3t$ at $U\gtrsim U_{c2}$.

For the  $U=12t \gtrsim U_{c2}$ Hubbard model on the  square
lattice, we found a half gap size of $2.1t$ in the PM phase and
$3.8t$ in AFM phase. Thus for $U\gtrsim U_{c2}$ antiferromagnetic
order increases the gap relative to that of  the paramagnetic
solution by about $80\%$. This has been qualitatively noted in
Ref.~\onlinecite{Sangiovanni06} and Ref.~\onlinecite{Comanac08}
which applied ED  and QMC respectively to a model with a
semicircular density of states, but our method provides a reliable
quantitative result for the square lattice model. Our finding
supports the conclusions of Ref.~\onlinecite{Comanac08} that $U$
must be somewhat less than the Mott critical value in the cuprates.
The difference of gap size is also apparent in the calculated
optical conductivity, in particular the AFM optical conductivity is
remarkably sharp near the gap edge, and has a corresponding sharp
feature at the upper edge of the upper Hubbard band. We believe this
is special to the square lattice, arising from the perfect nesting.

\section*{Acknowledgements}

We thank D. Reichman for helpful
discussions. XW and AJM are supported by NSF-DMR-0705847, EG by
NSF-DMR-0705847 and the Swiss National Science Foundation, LdM by
RTRA Triangle de la physique and MC by MIUR PRIN2007 2007FW3MJX.
Some of the calculations have been done using the ALPS
library.\cite{ALPS}

\end{document}